\documentclass{epl}
\usepackage{graphicx}
\usepackage[latin1]{inputenc}
\usepackage{pst-node,graphics,array}
\usepackage{dcolumn}
\usepackage{psfrag}
\title{A Modified Coupled-Cluster Based Investigation of Delocalization in Tight-Binding Hamiltonians}
\shorttitle{Delocalization Investigation}
\author{A. Boullanger\inst{1}  \and V. Robert\inst{1}}
\institute{
  \inst{1} Laboratoire de Chimie, Ecole normale sup\'erieure de Lyon,
 46 all\'ee d'Italie, 69364 Lyon Cedex 07, FRANCE \\
}

\begin{document}

\maketitle

\begin{abstract}
The nature of delocalization in a 1D system ruled by a  tight-binding 
Hamiltonian is investigated. Using a local evaluation of the ground
state energy, it is shown that the range of the delocalization effects is
rather limited. The method is applied to $1/2$ and $1/4$ filled systems.
\end{abstract}



\section{Introduction}
For the last few decades, physicists and chemists have put a lot of
effort into understanding the properties of low-dimensional
compounds~\cite{Day, Keller,Canadell}.  A major achievement was the
elucidation of the physical properties of polyacetylene the simplest
linear polymer~\cite{SSH}. Since then, a large number of inorganic
chains have been subject to extensive studies.  
Usually the chemically relevant and extremely simple model
Hamiltonian, the H\"uckel Hamiltonian, has been called
for~\cite{Salem, Hoffmann, ABW}. The use of more sophisticated
Hamiltonians such as the Hubbard Hamiltonian led to fundamental issues
since exact solutions are known for one dimensional systems
only~\cite{Wu-Lieb}.  
The traditional  solid-state picture gives a privilege to delocalization, presented
as a collective phenomenon.  However, it is known from the solid state
and the surface bonding that the local interactions ({\it i.e}. bonds)
play a dominant role and contain a large part of the chemical
actions~\cite{Hoffmann,Zintl}.
Following this trend, we
have recently demonstrated that a localized description of the physics
of a system ruled by very short-range interactions ({\it i.e.} Hückel
Hamiltonian) is relevant~\cite{Malrieu, Robert}. Indeed, starting from
strongly localized $N$-electron functions built from either pure
Atomic Orbitals (AOs) or fully localized bond Molecular Orbitals
(MOs), it is possible to get very good evaluations of the ground state
energy of a periodic lattice without any computational effort.  The
method consists in a self-consistent (SC) perturbation of a zeroth-order
wave-function which incorporates high order effects and offers
particular convergence properties.  
The charge transfer contributions are explicitly evaluated, a
noticeable difference with the original moments method~\cite{Moments}
which focuses on the charge densities.  
This modified coupled-cluster approach has shed the light on the crucial importance
of the local effects which appear to be the quantitatively dominating
driving forces responsible for the bond alternation in $1/2$ filled
conjugated hydrocarbons (i.e. Peierls distortion~\cite{Peierls}).  The
short and medium-range character of delocalization in the ground state
clearly appeared to be dominant. However, our evaluations based either
on pure AOs or bond MOs did not allow one to perfectly reproduce the
reference tight-binding calculations.  Thus, to which extent
delocalization has to be incorporated remains a challenging issue in
the study of delocalized systems.  Besides, the accurate determination
of the low-energy physics of such systems by rather local methods is
tempting since it opens the possiblity to determine the ground state
energy of systems ruled by a variety of model Hamiltonians with an
efficient and unexpensive method.

In this paper, we shall first briefly recall the method which has been
recently proposed by introducing the so-called set of self-consistent
coupled equations ({\em SCPEs})~\cite{Malrieu}. We will then follow
two different strategies based on this method.  The first one consists
in using MOs delocalized over four-site units, {\it i.e.} butadiene
MOs, to build the zeroth-order wave-function. The second aims
to extend the variational space which had been limited to single
charge transfer amplitudes~\cite{Malrieu,Robert}. Higher order
terms were incoporated by means of perturbation theory. Thus,
starting from a zeroth-order wave-function based either on AOs or MOs, single
and double charge transfers as well as second-nearest neighbour charge
transfers coefficients will be treated self-consistently.
For a strictly non-interacting electron system ({\it i.e.}
Hückel Hamiltonian), our goal is i) to look into the benefit of both
approaches in the evaluation of the ground state energy of a
$1/2$-filled homogeneous chain, ii) to extract an estimation of the
relevant delocalization range, and iii) to investigate the reliability
of such local evaluations for $1/4$-filled systems.
We believe that in the study of periodic lattices, local pictures
are very valuable since the information they convey is 
very instructive in the undertanding of bond formation analysis.

\section{Self-Consistent Perturbation Method}
Let us consider the H\"{u}ckel Hamiltonian $H = \sum
t(a^{\dagger}_{2i-1} a_{2i} + a^{\dagger}_{2i} a_{2i-1})$ ruling an
infinite chain of singly occupied valence orbitals.  The
nearest-neighbour interaction is determined by the hopping integral
$t$. The derivation of the H\"uckel energy for such $1/2$-filled
system is very standard and leads to $4/\pi \approx 1.273$ in the unit
of $t$~\cite{ABW}.  The basic idea behind the {\em SCPEs} method is to set the
electrons in strongly localized monoelectronic functions $\varphi_{i}$
to build a $N$-electron zeroth-order determinant $\Phi_0$.  Since the
$\alpha$ and $\beta$ electrons move independently, one may pair the
electrons on the odd sites to give rise to an Atomic Charge Density
Wave ({\em ACDW}) $\Phi_0 = \left| \varphi_{2i-1} \bar{\varphi}_{2i-1}
\right|$.  One may as well start from the bond MOs to describe a zeroth-order
Bond Density Wave ({\em BDW}) determinant.  Using the so-called
intermediate normalization, the wave-function is expanded in terms of
the charge transfer determinants $\{ \Phi_{k} \}$,
\begin{eqnarray}
\Psi = \Phi_0 + \sum_{k \neq 0} c_{k} \Phi_{k}
\end{eqnarray}
As in Ref.~\cite{Malrieu}, we define the first generation subspace
$S_{1}$ as the determinants $\{ \Phi_{k} \}_{k \neq 0}$
which interact with $\Phi_{0}$ 
({\it i.e.}  $\langle \Phi_0|H|\Phi_k \rangle \neq 0 $). 
Such determinants $\{ \Phi_{k} \}_{k \neq 0}$ are obtained
through charge transfers $T^{\dagger}_{k}$ between adjacent units. One
can easily show that the cohesive energy $E_{coh}$ is completely determined by
the knowledge of the $S_{1}$ determinants amplitudes, $E_{coh} = H_{00} +
\sum_{l \neq 0} c_{l} H_{0l}$. 
Let us briefly recall the derivation of the {\em SCPEs}
starting from the eigen-equation for the  $\Phi_{k}$
determinant :
\begin{eqnarray}
\left(H_{kk} - E \right)c_{k} + H_{k0} +   \sum_{l \neq k} H_{k l}c_{l}    +  \sum_{\alpha} H_{k \alpha}c_{\alpha} = 0 
\label{eq:eigen}
\end{eqnarray}
where the determinants $\alpha$ belong to the higher generation
subspaces.  If the $\Psi$ expansion is limited to
$S_{2}$-determinants, the $\{ c_{\alpha} \}$ coefficients are
determined by means of second-order perturbation theory.  The
$\Phi_{\alpha}$ can  be generated from $ \Phi_{k} $ by either the
action of a similar $T^{\dagger}_{l}$ operation or an
operation $R^{\dagger}_{m}$ which does not act on $\Phi_{0}$.  The
amplitude of the former may depend not only on the $c_{k} c_{l}$ product
associated with the $T^{\dagger}_{l} T^{\dagger}_{k}$ parent action on
$\Phi_0$ but also on the $c_{m} c_{n}$ products such that
$T^{\dagger}_{l} T^{\dagger}_{k}  = T^{\dagger}_{n} T^{\dagger}_{m}$. 
Let us mention that a major difference with the traditional coupled-cluster
approach~\cite{CC} lies in the  participation of the  $R^{\dagger}_{m}$ operators,
the role of wich has been recently investigated by means of a block-correlated   
coupled-cluster formalism~\cite{Shushua}. 
The derivation of the second-order amplitudes has been  given previously~\cite{Malrieu} and leads to :
\begin{eqnarray}
c_{\alpha} = c_{k} c_{l} \frac{\Delta E'_{k} + \Delta E'_{l}}{ \Delta E'_{k+l}}
\; +  \sum_{\scriptstyle \langle m,n \rangle \neq \langle k,l \rangle   
\atop  \scriptstyle T^{\dagger}_{m} T^{\dagger}_{n} = T^{\dagger}_{k} T^{\dagger}_{l}  }  
 \!\!\!\!\!\!\! c_{m}c_{n} \frac{\Delta  \! E'_{m} + \Delta  \! E'_{n} }{\Delta  \! E'_{k+l} }
\label{eq:xnadd}
\end{eqnarray}
where the excitation energies  $\Delta E_{k} =   H_{kk} - H_{00}$ are shifted by the 
{\em EPV} ({\em Exclusion Principle Violating}~\cite{Lindgren}) correction to deal with 
higher order corrections~\cite{Dressing} :
\begin{eqnarray}
\Delta  \!  E'_{k} = \Delta E_{k} - EPV(k) 
=  \Delta E_{k} \;\; - \sum_{l, T^{\dagger}_{l} \Phi_{k} = 0} \! \! \! \!  H_{0l}c_{l}
\end{eqnarray}
Let us mention that for $\Phi_{\alpha}$ determinants being generated by
a single pair  $T^{\dagger}_{l} T^{\dagger}_{k}$ the additivity in the excitation energies 
reduces $c_{\alpha}$ to the simple product $c_{k} c_{l}$, resulting in a perfect cancelation of
the   $c_{k} c_{l} H_{0l}$ term in eq.~\ref{eq:eigen}. Such situation is likely to
occur for remote excitations $T^{\dagger}_{k}$ and $ T^{\dagger}_{l}$, as expected
from simple physical considerations. 

The $R^{\dagger}_{m} T^{\dagger}_{k}  \Phi_{0}$ determinants amplitudes can be
immediately evaluated from second-order perturbation theory : 
\begin{eqnarray}
c_{\beta} = \sum_{n} \!   -\frac{H_{\beta n}}{\Delta  \! E'_{\beta}} c_{n}
\label{eq:xspecial}
\end{eqnarray}
Finally, the {\em SCPEs} in terms of the $\{ c_{k} \}_{k \neq 0}$ are derived
from eq.~\ref{eq:eigen} by expliciting the coefficients $c_{\alpha}$
using eqs.~\ref{eq:xnadd} and~\ref{eq:xspecial}. Let us mention that the dimension of the set of
coupled equations is equal to the number of charge transfer types in the system.

Part of the present work consists in a SC evaluation of the previous $c_{\alpha}$
and  $c_{\beta}$ coefficients.

\section{Extension of the Method}
For a homogeneous periodic lattice, the
cohesive energy $E_{coh}$ has been previously evaluated with absolute
relative errors $2.6\%$ and $1.5\%$ using AOs or bond MOs,
respectively~\cite{Malrieu}. Thus, it was rather challenging to start with larger units
built on the butadiene MOs (four-site MOs) to investigate the importance of
the collective effects.  In order to clarify the presentation, the  
strategy  will be  referred to as  $l, L$ where
$l$ stands for the size of the building blocks ($1$, $2$ or $4$) and $L$
for the extent of the SC treatment. $L=II$ indicates that the 
$S_{1}$ and $S_{2}$ non-factorizable determinants amplitudes are treated
in SC way.
   
\begin{figure}[htbp]
\begin{center}
\psfrag{h}{$h$}
\psfrag{i}{$i$}
\psfrag{j}{$j$}
\psfrag{k}{$k$}
\psfrag{l}{$l$}
\psfrag{a}{(a)}
\psfrag{b}{(b)}
\psfrag{c}{(c)}
\includegraphics[scale=0.6]{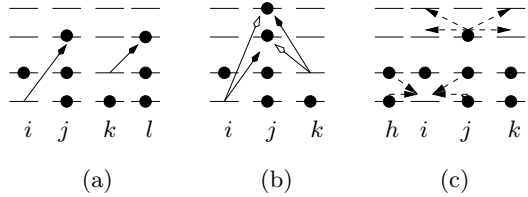} \\
\caption{Second-order determinants : (a) $T^{\dagger}_{k_2l_3}
T^{\dagger}_{i_1j_3}$, (b) with two parents, $T^{\dagger}_{k_2j_4}
T^{\dagger}_{i_1j_3} = T^{\dagger}_{k_2j_3} T^{\dagger}_{i_1j_4}$,
(c) $R^{\dagger}_{m} T^{\dagger}_{i}$ : the dotted arrows indicate
the different possible propagations. The black circles stand for the $\alpha$
electrons.}
\label{fig:S2-space}
\end{center}
\end{figure}

In using four-site MOs, the amount of delocalization
in the zeroth-order wave-function $\Phi_0$  is enhanced as compared to bond
MOs. Thus, one may expect to improve the cohesive energy evaluation.
The derivation of the Hückel energies $\{ e_{n} \}_{n=1-4}$ and MOs
coefficients of butadiene is very standard~\cite{ABW}.  Let us write
$\{ {i}_{n} \}_{n=1-4}$ the MOs localized on the $i^{th}$ tetrameric
unit associated with energies $\{ e_{n} \}_{n=1-4}$.  The energy per
atom corresponding to the zeroth-order determinant $\Phi_0 = \left|
{i}_{1} \bar{i}_{1} {i}_{2} \bar{i}_{2} \right|$ reads $E_{0}/t = (2
e_{1} + 2 e_{2})/4t =\sqrt5/2 \approx 1.118$. Starting from $\Phi_0$,
one can generate the $S_{1}$-subspace.
Since the left and
right charge transfer absolute amplitudes between adjacent tetrameric
units are equal, only four types are to be considered,
namely $c_{1} = c_{i_{1} \rightarrow j_{3}}$,
$c_{2} = c_{i_{1} \rightarrow j_{4}}$,
$c'_{1} = c_{i_{2} \rightarrow j_{3}}$ and 
$c'_{2} = c_{i_{2} \rightarrow j_{4}}$ .  As mentionned before, the $\alpha$
and $\beta$ electrons can be treated independently in the Hückel
framework ({\it i.e.} $E_{\alpha} = E_{\beta} = 1/2 E_{coh}$). 
Therefore, the cohesive energy reads :
\begin{eqnarray}
E_{coh}  = E_{0} +  (4 c_{1} t_{1} + 4 c_{2} t_{2} + 4 c'_{1} t'_{1} + 4 c'_{2} t'_{2} )/4
\label{eq:Etetramere}
\nonumber
\end{eqnarray}
Using the expressions of  the butadiene MOs in terms of the AOs, one immediately finds
the coupling matrix elements
$t_{1} = -t'_{2} = 0.224t$, $t'_{1} = -0.362t$ and $t_{2} = 0.138t$.   
The generation of the $S_{2}$-subspace is straightforward but somehow lengthy and
will not be fully detailed here. However, the three
types of determinants belonging to  $S_{2}$   are
shown in fig.~\ref{fig:S2-space}. 
The first one corresponds to a double charge transfer generated by a
single pair $\{ T^{\dagger}_{k},T^{\dagger}_{l} \}$. Conversely, the
second exemplifies the multiple generation of a double charge transfer
(see eq.~\ref{eq:xnadd}).  Finally, the third one gives rise to
contributions evaluated by eq.~\ref{eq:xspecial}.  The
derivation of the {\em SCPEs} for a $1/4$-filled system follows exactly the same
pattern by taking the inventory of the different $S_{1}$ and  $S_{2}$-subspaces determinants.
For both fillings, the determination of the nearest-neighbour charge transfer
amplitudes implies up to four butadiene units, that is 26 AOs, without eliminating
the translational property of the chain.

\begin{figure}[htbp]
\begin{center}
\psfrag{i}{$i$}
\psfrag{j}{$j$}
\psfrag{k}{$k$}
\psfrag{ph0}{$\Phi_{0}$}
\psfrag{phi}{$\Phi_{i}$}
\psfrag{phi+j}{$\Phi_{i+j}$}
\psfrag{phbeta}{$\Phi_{\beta}$}
\psfrag{phgama}{$\Phi_{\gamma}$}
\includegraphics[scale=0.5]{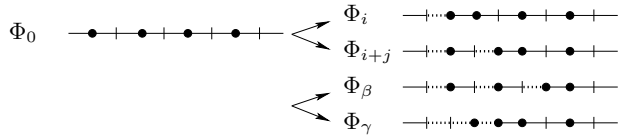} \\
\caption{Charge transfers generation using the AO based zeroth-order
determinant. Dotted lines indicate the electron propagations.}
\label{fig:S3-space}
\end{center}
\end{figure}

Instead of increasing the degrees of freedom one can extend the SC treatment.
Thus,  the second strategy we considered is based on an enlargement of the variational
space by treating the $S_{1}$ and non-factorizable $S_{2}$ determinants amplitudes $c$ and $d$ in a
SC way. Third-order terms are then evaluated from perturbation theory.
As a zeroth-order wave-function, we first used 
the strongly localized determinant built on the AOs.
The $S_{2}$-subspace consists of
the double charge transfers $\Phi_{i+j}$ which cannot be factorized as product of 
simple charge transfers (see fig.~\ref{fig:S3-space}).

The eigen-equation for the  $S_{1}$ determinants $\Phi_{i}$  is easily derived :
\begin{eqnarray}
 (-3ct)c + t + 2t(d-c^{2}) = 0
\label{eq:generation-c}
\end{eqnarray}
Starting from the  $S_{2}$ determinants  $\Phi_{i+j}$,
the adjacent excitation $T^{\dagger}_{k}$  gives rise to a
charge transfer  $\Phi_{\beta}$ (see fig.~\ref{fig:S3-space})
with amplitude $c_{\beta}$.
A third-order expansion gives :
\begin{eqnarray}
c_{\beta} = -d \left[ H_{k \beta}/ \Delta E'_{0 \beta} \right] = cd  \left[3ct / 7ct \right]
\nonumber
\end{eqnarray}
However, $\Phi_{\beta}$ can be generated by two other routes. The
first one which has the same amplitude $c_{\beta}$ results from the
action of $T^{\dagger}_{i}$ on $\Phi_{j+k}$.  One can also start from
the strictly additive double excitation
$T^{\dagger}_{k}T^{\dagger}_{i}$ ({\it i.e.} $c^{2}$ amplitude) and
then propagate an electron along the $j$ bond
through $T^{\dagger}_{j}$. The resulting contribution is $(3ct/7ct)
c^{3}$.
Then, one has to consider the action of $R^{\dagger}_{i}$ on
$\Phi_{i+j}$ (eq.~\ref{fig:S3-space}) which generates a third-order
determinant $\Phi_{\gamma}$ (see fig.~\ref{fig:S3-space}) with
amplitude $dt/4ct$ (eq.~\ref{eq:xspecial}).
Remembering that  $\Phi_{i+j}$ is directly coupled to the   $S_{1}$ determinants  $\Phi_{i}$
and   $\Phi_{j}$, the second eigen-equation reads :
\begin{eqnarray}
 (-5ct)d + 2ct + 2t \left[(3ct/7ct) (2 c d + c^{3}) -cd) \right] + 
\nonumber \\ t \left[d t/4 ct \right] = 0
\label{eq:generation-d}
\end{eqnarray}
whilst the cohesive energy $2ct$ is fully determined by $c$.  In the
following, eqs.~\ref{eq:generation-c} and \ref{eq:generation-d} are
referred to as {\em extended SCPEs}.

\begin{figure}[htbp]
\begin{center}
\psfrag{i}{$i$}
\psfrag{j}{$j$}
\psfrag{k}{$k$}
\psfrag{l}{$l$}
\psfrag{ph0}{$\Phi_{0}$}
\psfrag{phi+k}{$\Phi_{ij^{\star},kl^{\star}}$}
\psfrag{phik}{$\Phi_{ik^{\star}}$}
\psfrag{phgama}{$\Phi_{\gamma}$}
\includegraphics[scale=0.5]{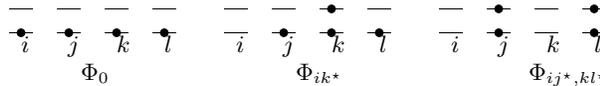} \\
\caption{Charge transfers generation using the bond MOs based zeroth-order
determinant.}
\label{fig:S2-MOs}
\end{center}
\end{figure}

In order to get a clearer understanding of the role of the
$R^{\dagger}_{m}$ operators, one can finally start from a bond MOs
zeroth-order wave-function $\Phi_{0}$ and use a similar second
generation logic to evaluate the cohesive energy $(1+c)t$.  It has
been previously observed that the deviation from non additivity is
negligible. Thus, the double charge transfer $\Phi_{ij^{\star},kl^{\star}}$ amplitudes
(see fig.~\ref{fig:S2-MOs}) can be approximated by $c^2$. However, the
$R^{\dagger}_{m}$ operators give rise to second-nearest neighbour
charge transfer $\Phi_{ik^{\star}}$ (see fig.~\ref{fig:S2-MOs}) the
amplitudes of which $d$ are self-consistently calculated.  Along this
framework, the third-nearest neighbour charge transfer
$\Phi_{il^{\star}}$ amplitudes are estimated from perturbation theory.
Such strategy combines the use of intermediate fragments size (bond
MOs) and an extension of the SC treatment. The {\em
extended SCPEs} are very easily derived in terms of the first
and second nearest-neighbour charge transfers $c$ and $d$, respectively.

For both strategies, the non-linear set of coupled equations is solved
without any computational effort. The cohesive energies for
$1/2$-filled systems are summarized in Table~\ref{table:Ecoh}. These
energies are compared to the ones of finite size clusters
$E_{coh}^{(n)}$, the extensions of which $n$ are measured by the
number of AOs implied in our approach.
%

\begin{table}[htbp]
\begin{center}
\begin{tabular}{ccccc}
\hline
\hline
Number of sites    &      &  1        &   2  &    4   \\
\hline  
\multicolumn{1}{c}{Extent of the}  & $I$   &   1.240$^{a}$ ( 1.205$^{(10)}$) &   1.253$^{a}$ (1.224$^{(14)}$) &  1.273  (1.246$^{(26)}$)  \\ 
\multicolumn{1}{c}{SC-treatment}  &  $II$   &   1.267 ( 1.216$^{(12)}$) &   1.272 (1.234$^{(18)}$) &   \\     &  &   \\
\hline
\hline
\end{tabular}
\caption{Cohesive energy evaluations for a $1/2$-filled system ($E_{exact} = 1.273$). The exponents indicate
the number of AOs used in the finite size clusters calculations given in parenthesis.$^{a}$Ref.~\cite{Malrieu}.  }
\label{table:Ecoh}             
\end{center}
\end{table}    
First, using the four-site based determinant, our local evaluation is
in excellent agreement with the reference value since it deviates by
less than $0.02\%$.  The {\em extended SCPEs} strategy starting from
the AO determinant (i.e. $(l,L)=(1,II)$) leads to a poorer , still satisfactory, estimation of
the cohesive energy. One should note that the delocalization effects
are smaller since the local evaluation is limited to $6$ lattice
parameters. Conversely, the combination of fragment of intermediate size (bond
MOs) and an extended SC treatment (i.e. $(l,L)=(2,II)$)  displays
excellent agreement with the exact value and rather inexpensive cost.
Therefore, considering the perturbative treatment which incorporates
high-order effects through the {\em EPV} corrections, we may conclude
that the delocalization effects are effective up to roughly $10$
lattice parameters.  The benefit of our approach which explicitely
includes the translational symmetry is significant as seen from the
finite size clusters calculations (Table~\ref{table:Ecoh}).  Second,
inspection of the different contributions to the cohesive energy shows
that a major part can be attributed to the $R^{\dagger}_{m}$
excitations which are not active in a strict coupled-cluster
formalism. Indeed, if one sets these terms to zero in the {\em SCPEs},
the relative error deteriorates ($0.02\% \rightarrow 1.6\%$). The
comparison between the {\em SCPEs} $(2,I)$ and the {\em extended
SCPEs} $(1,II)$ strategies is very instructive. Indeed, the numbers of
AOs implied are very similar. However, the use of a larger variational
space in the {\em extended SCPEs} noticeably enhances the accuracy
($1.5\% \rightarrow 0.5\%$). Interestingly, the $d/c^{2}$ ratio is
$\approx 1.253$ whereas based on the bond MOs, the deviation from
additivity of the local charge transfers is measured by the {\em EPV}
corrected excitation energies ratio, that is $(3ct + 3ct)/5ct = 1.2$~\cite{Malrieu}.
Thus, using the {\em extended SCPEs} the interference between local
excitations is somehow increased and a better estimation
results. Similarly, inspection of the second-nearest neighbor charge
transfer amplitudes shows that their contribution is significantly
enhanced ($d=0.155$) as compared to the original first generation
approach ($d=2c (t/2)/(2t + 3ct/2) = c/(2+3/2c) =
0.106$)~\cite{Malrieu}.

A very similar approach was applied to a $1/4$-filled system. Starting
from two-site or four-site MOs, the local evaluations deviate from the exact
value ($0.900$) by $2.9\%$ ($0.874$) and $0.55\%$ ($0.895$),
respectively. Even though a relative agreement with the reference
exact value is reached , the errors are somehow larger. From band
structure calculations, it is known that the higher the energy in the
dispersion curve, the more numerous the nodes in the orbital. Thus, we
may argue that $1/4$-filled systems are more delocalized than
$1/2$-filled ones.  The accuracy of our approach which introduces only
short-range delocalization effects clearly follows the delocalization
nature of the system. However, the {\em extended SCPEs} method still
leads to a more accurate estimation of the $1/4$-filled system energy
($0.895$) than the one obtained from a $26$ AO cluster ($0.831$).

\section{Conclusions}
In summary, the method we used allowed us to investigate the delocalization
effects in a 1D-chain. We showed that the latter are limited since
excellent evaluations for both $1/4$ and $1/2$ filled systems can be
obtained from rather local evaluations.  Besides, the method offers a
way to estimate the delocalization range by  step by step incorporating
the short and medium range contributions. Beyond roughly a $10$-lattice
parameter distance,  their importance is vanishingly small.
From a conceptual point a
view, the original {\em SCPEs} approach~\cite{Malrieu} can be developed at will by
including higher order terms in the variational space. Such strategy
is very powerful since 
the benefit in terms of accuracy is significant whilst
the  {\em SCPEs} are easily derived.
In the view of our results, we believe that the  description of important
heteroatomic $\left( AA' \right)_{n}$ chains which are more localized
by essence would not be of great interest since the bond MOs description
has already produced very satisfactory estimations~\cite{Malrieu}. 
However, the methodology
can be easily extended to more sophisticated Hamiltonians, including
for instance one-site $U$ and inter-site $V$ repulsions, the 
resolutions of which are traditionally derived from expensive approaches.

\acknowledgments      
The authors are very grateful to J.-P Malrieu for enlightening
discussions. 



\end{document}